\documentclass[preprint]{aastex6}
\begin{document}
\title{New or Improved Orbital Periods of Cataclysmic Binaries}

% have to somehow get a line skip into the author list 
\author{John R. Thorstensen}
%\affil{Dartmouth College Department of Physics and Astronomy\\
%6127 Wilder Laboratory\\
%Hanover, NH 03755-3528}
\author{Frederick A. Ringwald}
%\affil{Dartmouth College Department of Physics and Astronomy\\
%6127 Wilder Laboratory\\
%Hanover, NH 03755-3528}
\author{Cynthia J. Taylor}
%\affil{Dartmouth College Department of Physics and Astronomy\\
%6127 Wilder Laboratory\\
%Hanover, NH 03755-3528}
\author{Holly A. Sheets}
%\affil{Dartmouth College Department of Physics and Astronomy\\
%6127 Wilder Laboratory\\
%Hanover, NH 03755-3528}
\author{Christopher S. Peters}
%\affil{Dartmouth College Department of Physics and Astronomy\\
%6127 Wilder Laboratory\\
%Hanover, NH 03755-3528}
\author{Julie N. Skinner}
%\affil{Dartmouth College Department of Physics and Astronomy\\
%6127 Wilder Laboratory\\
%Hanover, NH 03755-3528}
\author{Erek H. Alper}
%\affil{Dartmouth College Department of Physics and Astronomy\\
%6127 Wilder Laboratory\\
%Hanover, NH 03755-3528}
\author{Kathryn E. Weil}
\affil{Dartmouth College Department of Physics and Astronomy\\
6127 Wilder Laboratory\\
Hanover, NH 03755-3528}

\section{Explanation}

Cataclysmic variables (CVs) are close binary star systems in which a white
dwarf accretes mass from a low-mass companion via Roche-lobe overflow.  
%The name arises from the dramatic photometric variations,
%most commonly dwarf-nova outbursts, that many such systems show.
\citet{warner95} presents a comprehensive review.

The most significant observable number for a CV is
its orbital period $P_{\rm orb}$.
%, which correlates with its
%morphology and indicates its evolutionary status.  
%Many
%CV orbital periods are measured during outburst by amateur 
%astronomers, usually from the superhumps exhibited
%by the SU UMa subtype, from which $P_{\rm orb}$ can
%be estimated to $\sim 1$ per cent. However, many other objects
%are less tractable.  Most non-eclipsing dwarf novae, for
%example, reveal their periods through radial velocity time 
%series taken near minimum light.
We report here CV orbital periods measured over the last 
several decades using the Hiltner 2.4m and McGraw-Hill 1.3m 
telescopes at MDM Observatory on Kitt Peak, Arizona, mostly with 
radial-velocity spectroscopy. While most of our results
have been published in peer-reviewed articles (see, 
e.g., \citealt{thorzoo,thorskinner,thorsdss,thorlongp}), 
%but preparing detailed discussions incurs delay, and  
many periods we have determined remain unpublished. 
%In addition, the advent of automated, large-scale synoptic sky
%surveys such as the Catalina surveys \citep{drake14,
%drakevars14}, ASAS-SN \citep{shappeecurtain}, 
%and others has led to a flood of new CV discoveries. 
%We have determined $P_{\rm orb}$s for many of the most
%tractable new CVs; while this is useful, it tends to exacerbate 
%the backlog.

To make our CV periods available more quickly, and
prevent duplication of effort, we 
%thought it worthwhile to 
present in Table 1 some of our heretofore unpublished $P_{\rm orb}$s, 
%\footnote{We are including primarily systems 
%unencumbered by co-authorships.}, reserving details for forthcoming publications.
some of which are refinements of previously-known periods. 
% based on longer time bases. 
The instrumentation, observing protocols, reduction, and analysis are 
described in the papers cited above.  %We are 
%reasonably confident that the periods are based on correct choices of 
%cycle count -- mostly daily cycle count, but also run-to-run cycle count for
%periods quoted to higher precision.  
We are confident of the choices of cycle count between observations.
% and the 
%quoted uncertainties.

%The nomenclature used for stars without variable star names
%is heterodox; in cases of doubt, 
%The coordinates given establish a unique identification.
For references on these stars, see
SIMBAD and the online lists maintained
by ASAS-SN, the Catalina surveys, and MASTER.
\citet{downes01} give references for long-known CVs 
(see also \citealt{downes05}).

\newpage

%Those seeking additional detail on any of these objects 
%Please contact the lead author for additional detail.
%are urged to contact the lead author.

% \startlongtable
\begin{deluxetable}{llllll}
\tablecolumns{6}
%\tabletypesize{\scriptsize}
\tabletypesize{\footnotesize}
\tablewidth{0pt}
\tablecaption{New or Improved Orbital periods}
\tablehead{
\colhead{Name} &
\colhead{$\alpha_{\rm 2000}$} &
\colhead{$\delta_{\rm 2000}$} & 
\colhead{$P$} &
\colhead{$T_0$\tablenotemark{a}} & 
\colhead{Method\tablenotemark{b}} \\
\colhead{} &
\colhead{[h m s]} &
\colhead{[$^{\circ}$ $'$ $''$]} &
\colhead{[d]} &
\colhead{} &
\colhead{} \\
}
\startdata 
ASAS-SN 16lu             & 00 24 37.16 &   +11 06 11.2   & 0.3302(1) & & abs,emn \\
FL Psc                   & 00 25 11.09 &   +12 17 12.3   & 0.0560(1) & & emn \\
1RXS J012750.5+380830    & 01 27 50.60 &   +38 08 12.0   & 0.06071(5) & & emn  \\
RX J0153.3+7446          & 01 53 20.94 &   +74 46 21.7   & 0.16415(1) & & emn  \\
WY Tri                   & 02 25 00.47 &   +32 59 55.8   & 0.07569(7) & & emn  \\
BB Ari                   & 02 44 57.78 &   +27 31 09.2   & 0.0702(1)  & & emn  \\
CSS071108:030945+295251  & 03 09 45.19 &   +29 52 50.9   & 0.08468(3) & & ecl,emn \\
SDSSJ032015.29+441059.3  & 03 20 15.29 &   +44 10 59.3   & 0.0687(12) & & emn \\
V1294 Tau                & 04 00 37.25 &   +06 22 46.2   & 0.1490188(2) & 53017.792(4) & emn  \\
V1024 Per                & 04 02 39.02 &   +42 50 46.0   & 0.0706(3) & & emn \\
V1389 Tau                & 04 06 59.78 &   +00 52 44.3   & 0.07815(6) & & emn \\
ASAS-SN 16pm             & 04 17 24.77 &   +22 15 22.8   & 0.1540(2) & & emn \\
FY Per                   & 04 41 56.58 &   +50 42 36.2   & 0.2584(3) & & emn \\
V1208 Tau                & 04 59 44.03 &   +19 26 22.9   & 0.0681(2) & & emn \\
V1193 Ori                & 05 16 26.65 &   $-$00 12 14.2   & 0.1657(3) & & emn \\
Paloma                   & 05 24 30.52 &   +42 44 50.4   & 0.1092(3) & & emn \\  
1RXS J055722.9+683219    & 05 57 18.47 &   +68 32 27.0   & 0.0523(3) & & emn \\
V405 Aur                 & 05 57 59.29 &   +53 53 44.9   & 0.1726255(3) & 51838.048(5) & emn  \\
LS Cam                   & 05 57 23.96 &   +72 41 52.4   & 0.1423853(5) & 53743.942(4) & emn  \\
ASAS-SN 15ex             & 07 20 54.19 &   +34 12 45.3   & 0.16581(2) & & abs \\
% KQ Mon                   & 07 31 21.09 &   $-$10 21 50.0 & & &   \\
ASAS-SN 15tc             & 07 56 00.68 &   $-$02 51 12.3 & 0.1651(4) & & emn \\
SBSS 0755+600            & 07 59 26.39 &   +59 53 51.1   & 0.07334(8) & & emn \\
HZ Pup                   & 08 03 22.80 &   $-$28 28 28.8 & 0.212(1) & & emn  \\
SU UMa                   & 08 12 28.2  &   +62 36 22.6   & 0.07637540(8) & 50839.911(3) & emn \\
ASAS-SN 16as             & 09 14 10.75 &   +01 37 32.9   & 0.2518(3) & & emn,abs \\
% 1RXS J092737.4$-$191529  & 09 27 37.10 &   $-$19 15 34.0 & & &   \\
RW Sex                   & 10 19 56.62 &   $-$08 41 56.0 & 0.2451450(6) & 52233.922(4) & emn  \\ 
LY UMa                   & 10 48 17.98 &   +52 18 30.0   & 0.2712788(2) & 51994.741(1) & abs  \\
QZ Vir                   & 11 38 26.96 &   +03 22 08.1   & 0.05882047(7) & 53448.827(2) & emn \\
SDSS J134441.8+204408    & 13 44 41.83 &   +20 44 08.3   & 0.070592(4) & & emn  \\
TT Boo                   & 14 57 44.76 &   +40 43 40.8   & 0.075940(3) & & emn  \\
QZ Lib                   & 15 36 16.00 &   $-$08 39 07.6 & 0.06413(8) & & emn \\
SDSS J154453.6+255348    & 15 44 53.6  &   +25 53 48.8   & 0.25128168(2) & 54658.6372(2) & ecl \\
CSS170517:155156+145333  & 15 51 55.62 &   +14 53 32.9   & 0.0697(2) & & emn \\
QZ Ser                   & 15 56 54.46 &   +21 07 19.8   & 0.08316078(7) & 52438.8144(4) & abs \\
SDSS J155720.75+180720.2 & 15 57 20.75 &   +18 07 20.2   & 0.0810(3) & & emn \\
CSS160906:160346+193540  & 16 03 46.08 &   +19 35 39.9   & 0.05693(2) & & emn \\
ASAS-SN 15kw             & 17 19 37.10 &   +04 31 23.8   & 0.0592(2) & & emn \\
V380 Oph                 & 17 50 13.66 &   +06 05 29.1   & 0.1534766(4) & 53549.845(4) & emn \\
IX Dra                   & 18 12 31.43 &   +67 04 46.1   & 0.0648(2) & & emn \\
% complex alias structure in V868 Cyg summarized by this error bar.
V868 Cyg                 & 19 29 04.41 &   +28 54 25.4   & 0.1698(15) & & emn \\
1RXS J192926.6+202038    & 19 29 27.82 &   +20 20 35.4   & 0.162(1) & & emn \\
V2289 Cyg                & 19 34 36.14 &   +51 07 40.8   & 0.142(2) & & emn \\
RX J1946.2-0444          & 19 46 16.36 &   $-$04 44 56.8 & 0.07462(7) & & emn \\
UU Aql                   & 19 57 18.68 &   $-$09 19 20.8 & 0.1635324(3) & 48722.951(4) & emn  \\
V2306 Cyg                & 19 58 14.50 &   +32 32 41.9   & 0.1821454(8) & 53622.783(6) & emn \\
V1316 Cyg                & 20 12 13.65 &   +42 45 51.1   & 0.07412(15) & & emn \\
TW Vul                   & 20 39 34.48 &   +27 15 55.6   & 0.1591(6) & & emn \\
SDSS J204314.02+153602.0 & 20 43 14.02 &   +15 36 02.1   & 0.1515(3) & & emn  \\
ASAS-SN 15rz             & 20 53 05.42 &   +59 17 31.8   & 0.208(1) & & emn  \\
V1060 Cyg                & 21 07 42.20 &   +37 14 08.7   & 0.292(2) & & emn,abs \\
V444 Peg                 & 21 37 01.83 &   +07 14 45.9   & 0.0926(2) & & emn \\
KM Lac                   & 22 13 49.30 &   +55 28 27.0   & 0.21542426(2) & 57550.9018(7) & ecl \\
V368 Peg                 & 22 58 43.48 &   +11 09 11.9   & 0.0685(3) & & emn \\
V405 Peg                 & 23 09 49.17 &   +21 35 17.7   & 0.1776464(2) & 53182.019(2) & abs \\
EI Psc                   & 23 29 54.30 &   +06 28 10.9   & 0.044566904(6) & 52670.5745(2) & abs \\
NSV14652                 & 23 38 48.70 &   +28 19 55.5   & 0.07824(9) & & emn \\
ASAS-SN 15nv             & 23 42 26.24 &   +32 48 20.6   & 0.1348254(12) & 58040.6917(3) & ecl \\
V630 Cas                 & 23 48 51.91 &   +51 27 39.3   & 2.56388(2) & 50709.76(2) & abs \\
\enddata 

\tablenotetext{a} {Barycentric Julian date, minus 2 400 000., 
of either (a) the blue-to-red crossing through the mean velocity or (b) 
mid-eclipse, given only when the period is accurate enough
to predict phase in the present epoch.  The time system is UTC.}
% When both emission and absorption
%velocities are noted, the absorption value is given.}
\tablenotetext{b} {Measurements used for period finding; `em' and `abs' 
denote emission- and absorption-line radial velocity; `ecl' 
denotes eclipses.}
\end{deluxetable}

\end{document}